# Quantum Hall effect in epitaxial graphene with permanent magnets

F.D. Parmentier[1,*], T. Cazimajou[1], Y. Sekine[2], H. Hibino[2], H. Irie[2], D.C. Glattli[1], N. Kumada[2], and P. Roulleau[1]

[1]SPEC, CEA, CNRS, Université Paris-Saclay, CEA Saclay 91191 Gif-sur-Yvette cedex, France
[2]NTT Basic Research Laboratories, NTT Corporation,3-1 Morinosato-Wakamiya, Atsugi, Kanagawa, Japan
*corresponding author: francois.parmentier@cea.fr
(Dated: 28th September 2016)

## ABSTRACT

We have observed the well-kown quantum Hall effect (QHE) in epitaxial graphene grown on silicon carbide (SiC) by using, for the first time, only commercial NdFeB permanent magnets at low temperature. The relatively large and homogeneous magnetic field generated by the magnets, together with the high quality of the epitaxial graphene films, enables the formation of well-developed quantum Hall states at Landau level filling factors $\nu = \pm 2$, commonly observed with superconducting electro-magnets. Furthermore, the chirality of the QHE edge channels can be changed by a top gate. These results demonstrate that basic QHE physics are experimentally accessible in graphene for a fraction of the price of conventional setups using superconducting magnets, which greatly increases the potential of the QHE in graphene for research and applications.

## INTRODUCTION

Experimentally demonstrated for the first time in 1980 [1], the quantum Hall effect (QHE) is one of the most striking phenomena in condensed matter physics. It arises when a two-dimensional electron gas (2DEG) is immersed in a strong perpendicular magnetic field. In the quantum Hall state, the bulk of the 2DEG is insulating and the current flows in ballistic and chiral one-dimensional channels along the edges of the samples [2].

This very peculiar state of matter has been the object of an extremely vast corpus of fundamental research on various materials. Indeed, quantum Hall systems host a great variety of unique states and phenomena, such as fractal quantum Hall states [3,4,5,6]. Furthermore, the ballistic edge channels of the QHE make it possible to engineer devices implementing textbook concepts of mesoscopic quantum transport, such as the electronic Mach-Zehnder interferometer [7]. When coupled to a superconductor, QHE edge channels have recently been predicted to host propagating Majorana Fermions [8]. The QHE is currently used in metrology to define a practical resistance standard [9], and the non-dissipative nature of the QHE edge channels is extremely attractive for applications in low-power and high speed electronics.

Experimentally, the QHE is obtained in materials such as gallium arsenide heterostructures, under a strong magnetic field (of at least a few Tesla) and at low temperatures. These constraints, which require the use of cryogenic refrigerators containing superconducting magnets to generate the magnetic field, stem from the small energy gap between Landau levels at low magnetic fields [10].

Graphene can circumvent those constraints because, owing to its unique band structure, the energy gap is more than one order of magnitude larger than in gallium arsenide heterostructures at typical magnetic fields of a few Tesla.

Indeed, shortly after the first experimental observation of the QHE in graphene [11], it has been demonstrated that graphene shows signatures of the QHE at room temperature when subjected to very large magnetic fields [12].

The ability to tune the density of carriers, and even their type, in graphene by the field effect implies that the constraint of high magnetic fields could also be lifted for clean enough samples, in which carrier densities lower than $5 \times 10^{10}$ cm$^{-2}$ can be attained, making the QHE at the Landau level filling factor $\nu = 2$ reachable for magnetic fields lower than 1 T [13,14]. Such fields are then compatible with superconducting materials such as niobium, enabling the coexistence of superconductivity and QHE edge channels [15,16]. Furthermore, magnetic fields of up to 1 T can now be easily reached using inexpensive rare earth permanent magnets, such as neodymium (NdFeB) magnets made of $Nd_2Fe_{14}B$ alloy, which have remanence fields of up to 1.4 T. Replacing the expensive, cumbersome and fragile superconducting solenoid magnets with affordable permanent ones would considerably expand efforts to probe and exploit the properties of the QHE.

In this paper, we show that the QHE can be induced in a graphene sample using NdFeB permanent magnets instead of the usual superconducting magnets.

Several fabrication techniques yield clean enough graphene samples so as to reach the QHE at sub-tesla magnetic fields. While boron nitride-encapsulated graphene [17] provides the highest mobilities and is particularly well suited for fundamental research, epitaxial graphene [18], grown either by thermal decomposition or by chemical vapor deposition on (6H-SiC(001)) silicon carbide (SiC) combines relatively high mobility, a very large single-crystal area, and high reproducibility, making it an excellent candidate for both research and applications. The QHE in epitaxial graphene has been extensively studied since its first observation [19], particularly the extremely robust filling factor $\nu = 2$ QHE plateau with respect to both gate voltage and magnetic field [20, 21, 22, 23]. This unique robustness makes epitaxial graphene a very appealing candidate for the realization of a practical resistance standard for metrology [24, 25, 26]. While all the aforementioned experiments were performed with superconducting magnets, we have used an epitaxial graphene sample installed in a custom-built sample holder, detailed below, to observe the QHE using only permanent magnets.

## RESULTS

Figure 1(A) is a schematic illustration of the sample holder. The holder is equipped with a commercial ceramic leadless chip carrier (LCC) socket through which a hole was bored so as to fit a cylindrical NdFeB permanent magnet (length ≈1 cm, diameter ≈6 mm) under the LCC. To generate an homogeneous magnetic field in a Helmholtz configuration, two additional magnets are aligned above the sample, separated from it by a ≈1 mm-thick spacer. As a result, the graphene sample sits in the middle of a ≈2 mm vertical gap, in which a magnetic field of ≈0.94 T (as measured using a room temperature Gaussmeter) is generated. A photograph of the system, as implemented in a dry He3 fridge, is shown in Fig. 1(B).

We used this setup to perform QHE measurements in an epitaxial graphene Hall bar. An optical micrograph of the sample is shown in Fig. 1(C). A top gate covering the Hall bar is used to tune the carrier density in the sample. Although the graphene films are intrinsically heavily doped (density ≈ $2 \times 10^{12}$ cm$^{-2}$), the top gate dielectric, made of hydrogen silsesquioxane (HSQ, see methods), decreases the carrier density to around $5 \times 10^{11}$ cm$^{-2}$ at zero top gate voltage $V_{gate}$, leading to high mobilities (up to 16000 cm$^2$/Vs at $V_{gate}$ =0 V). This, combined with the ability to further decrease the density by changing $V_{gate}$, allows us to reach the QHE regime at magnetic fields as low as 0.5 T, which is easily attainable with permanent magnets. This is clearly shown in Fig. 2, which presents the Hall ($R_H$) and longitudinal ($R_{xx}$) resistances under the magnetic field generated only by permanent magnets as a function of $V_{gate}$ at the base temperature of our He3 fridge (300 mK).

The results plotted in Fig. 2 show a quantization of $R_H$ (red squares) at the value $0.5 \times h/e^2$, accompanied by a zero value of $R_{xx}$ (black circles), for $V_{gate}$ between -33 and -39 V. This is the hallmark of the QHE at filling factor $\nu = 2$. As $V_{gate}$ is further swept towards negative values, the $\nu = 2$ plateau vanishes and the charge neutrality point (CNP) is reached at $V_{gate} \approx -45$ V. This is indicated by the local maximum in $R_{xx}$ and the change in the sign of $R_H$. For $V_{gate}$ between -50 and -55 V, $R_{xx}$ again reaches zero while $R_H$ becomes quantized at $-0.5 \times h/e^2$, characteristic of the QHE at filling factor $\nu = -2$. This change in sign, corresponding to a change in the chirality of charge carriers, is a signature of the ambipolarity of charge transport in graphene.

Having shown that our setup enables us to reach the QHE regime at $\nu = \pm 2$ in epitaxial graphene, we now focus on the robustness of the QHE phase against temperature. Figure 3(A) shows the temperature dependence of $R_H$ (red squares) and $R_{xx}$ (black circles) on the $\nu = +2$ plateau at up to T=2 K. A sevenfold increase in the temperature only leads to a deviation smaller than one-thousandth of the expected values for $R_H$. Note that the low-temperature value of $R_H$ is slightly smaller than $0.5 \times h/e^2$, which is due to an error in the estimated gain of the room-temperature amplifiers of about 0.04 %.

The deviations become larger as the temperature is increased, as shown in the inset of Fig. 3(B), where the $V_{gate}$ dependence of $R_{xx}$ is plotted for several values of temperature up to 30 K. Note that fluctuations arise in $R_{xx}$ around the CNP below 7 K, indicating the presence of residual electron/hole puddles at the CNP. In the explored temperature range, the minimum longitudinal conductivity $\sigma_{xx} = R_{xx}/(R_{xx}^2 + R_H^2)$, plotted in the main panel of Fig. 3(B), is reasonably well reproduced by a model combining thermal activation and Mott variable range hopping (VRH) [27]: $\sigma_{VRH} \propto 1/T \times \exp[(T_M/T)^{1/3}]$. This model is similar to the one presented in [23], used to reproduce the temperature dependence of the minima in $\sigma_{xx}$ in epitaxial graphene samples immersed in magnetic fields generated by a superconducting magnet.

## DISCUSSION

In summary, we have combined the low carrier densities attainable in epitaxial graphene with the relatively large magnetic fields generated by NdFeB permanent magnets to obtain a robust quantum Hall phase without the use of either superconducting or normal electro-magnet. In those conditions, the QHE could be reached in various types of graphene samples. For fundamental physics, using this setup with encapsulated monolayer and bilayer graphene [17], which shows the presence of the QHE for fields as low as 100 mT [28], would open the study of most of the basic features of the QHE to a very large audience. The lack of versatility of our technique with respect to superconducting magnets (*e.g.* one cannot reverse *in-situ* the direction of the magnetic field) can be mitigated by implementing a mechanical displacement of the magnets. In addition, it is possible to combine it with a small coil to induce small changes in the applied magnetic field [29]. For applications, the large scale and reproducibility of epitaxial graphene samples makes it the perfect material to use in conjunction with permanent magnets in order to engineer ambipolar, chiral and ballistic electronic devices, such as quantum Hall circulators [30, 31]. Finally, our results could open the way to compact, portable and practical metrology kits using epitaxial graphene to define the electrical resistance standard.

# MATERIALS AND METHODS

The epitaxial graphene wafers were prepared by thermal decomposition of a 6H-SiC(0001) substrate, annealed at around 1800 °C in Ar at a pressure of less than 100 Torr. Metallic contacts are made of Cr/Au. An Au top gate, separated from the graphene film by a 100-nm-thick hydrogen silsesquioxane (HSQ) and a 60-nm-thick $SiO_2$ insulating layers, allows us to tune the density and the type of the charge carriers. At zero gate voltage, the sample has n-type carriers with a density of about $5\times10^{11}$ $cm^2$, due to the doping from the SiC substrate and the HSQ. The measurements were performed using standard lock-in techniques at roughly 30 Hz with NF LI-75A room-temperature amplifiers.

**Acknowledgements:** The authors thank W. Escoffier and R. Ribeiro-Palau for fruitful discussions. This work was supported by the French ANR (ANR-11-NANO-0004 Metrograph) the ERC (ERC-2008-AdG MEQUANO) and the CEA (Projet phare ZeroPOVA). The authors declare no competing interests.

**Author contributions:** F.D.P. and T.C. performed the measurements and analysis.
Y.S., H.H. and H.I. grew the wafer and fabricated the samples, with inputs from N.K.
D.C.G., N.K., P.R. and F.D.P. supervised the project.
F.D.P. conceived the experiment and wrote the manuscript, with inputs from D.C.G., N.K., and P.R.

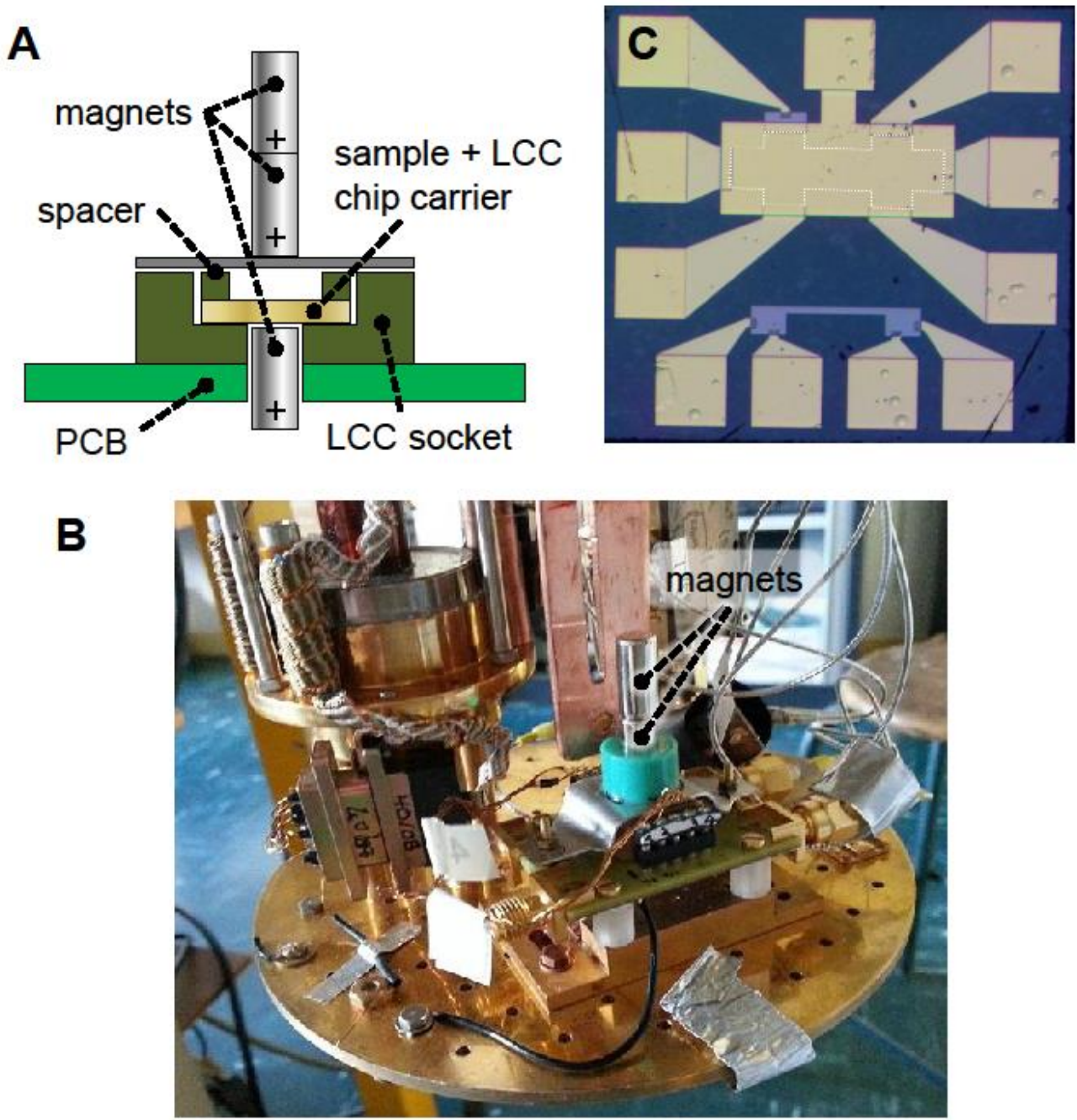

**Fig. 1. Experimental setup and sample. (A),** Schematic of the sample holder with three cylindrical NdFeB magnets generating an axial magnetic field of roughly 1 T in a Helmholtz configuration. The sample is placed between the lowermost magnet and the middle one. **(B),** Photograph of the setup implemented in our dry He3 fridge. **(C),** Optical micrograph of the graphene Hall bar sample, entirely covered by a top gate. The width of the Hall bar (indicated by the white dotted line) is 100 μm.

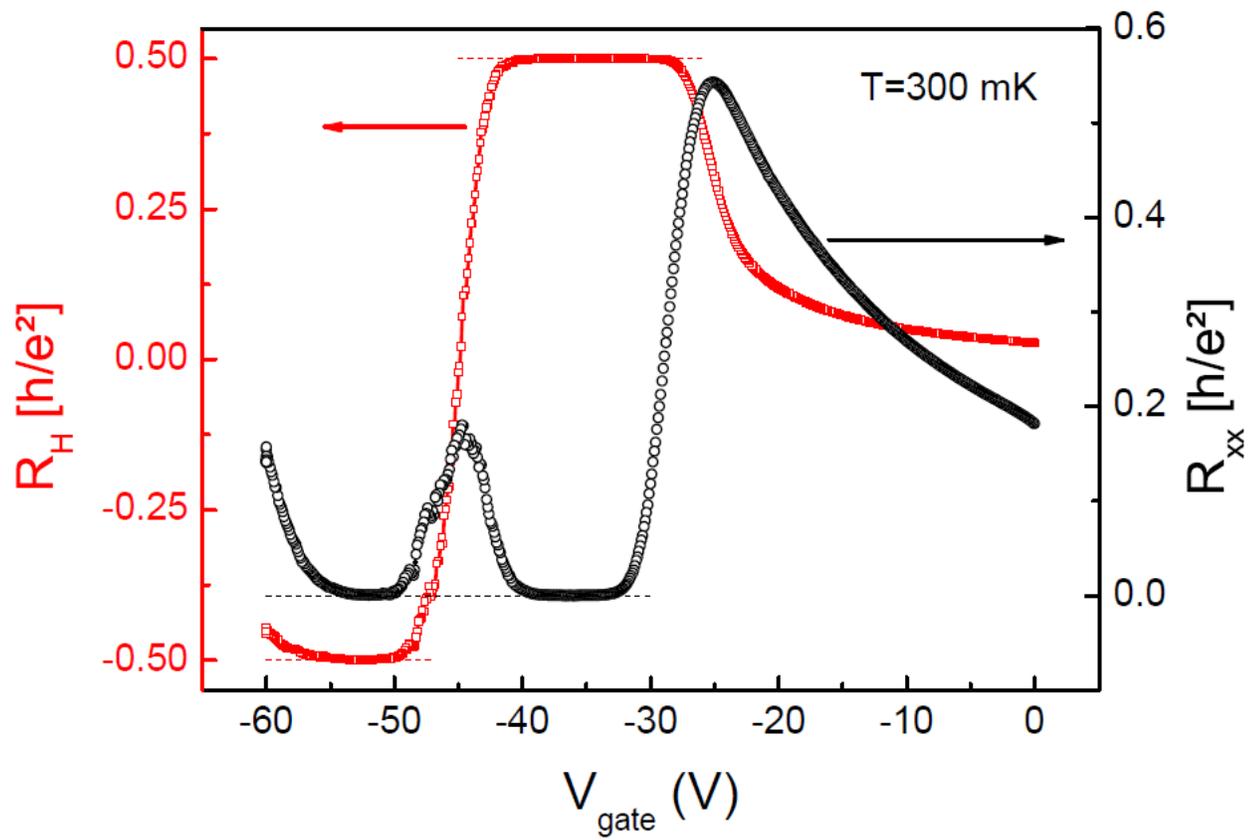

**Fig. 2. QHE with permanent magnets.** Hall resistance $R_H$ (red squares, left vertical axis) and longitudinal resistance $R_{xx}$ (black circles, right vertical axis) as a function of top gate voltage $V_{gate}$ measured at 300 mK. The thin dashed lines correspond to the expected values of $R_H$ (red) and $R_{xx}$ (black) for the QHE at filling factors $\nu = \pm 2$.

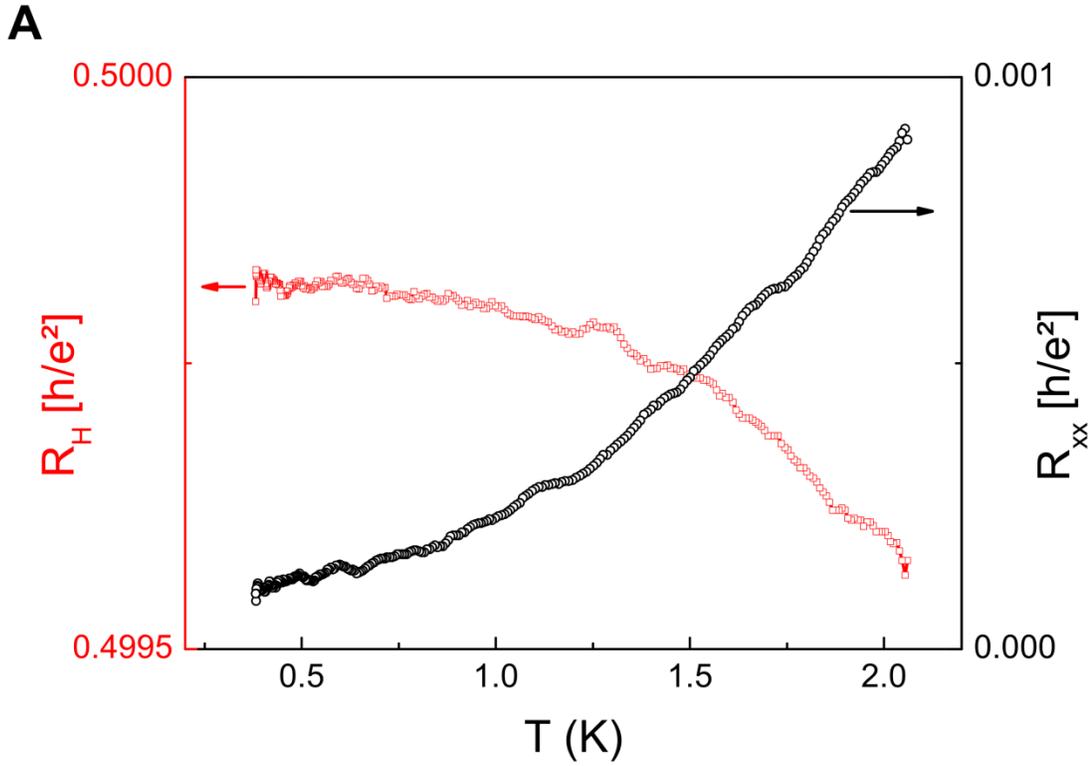

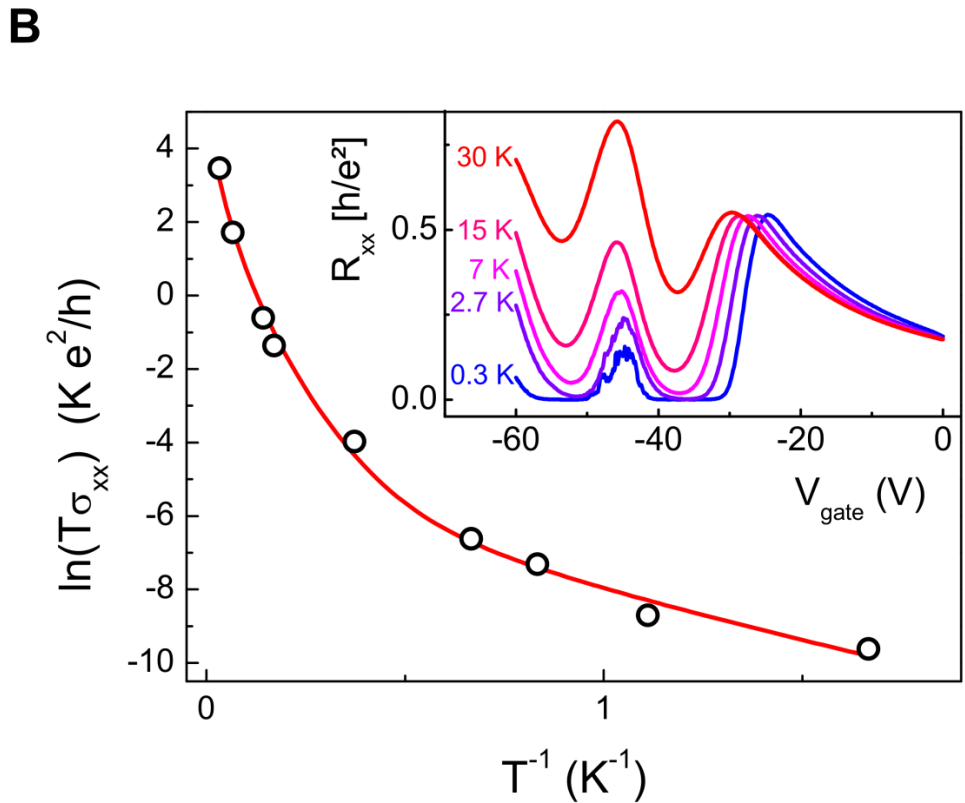

**Fig. 3. Temperature dependence. (A),** Temperature dependence of $R_H$ (red squares, left vertical axis) and $R_{xx}$ (black circle, right axis) on the $\nu=+2$ plateau at $V_{gate} =-38$ V, between 300 mK and 2 K. **(B),** Plot of $\ln(T\sigma_{xx})$ vs $T^{-1/2}$ (circles), where $\sigma_{xx}$ is measured on the $\nu=+2$ plateau, for temperatures between 300 mK and 30 K. The red line is a fit combining variable range hopping and thermally activated transport (see text). Inset: $R_{xx}$ as a function of $V_{gate}$ for several temperatures, up to 30 K.